\documentclass{aastex63}

\newcommand\webb{\textit{JWST}}
\newcommand\hubble{\textit{HST}}
\shorttitle{PACMan2: Proposal Review Management}
\shortauthors{Strolger et al.}
\graphicspath{{./}{figures/}}

\begin{document}

\title{PACMan2: Next Steps in Proposal Review Management}

\correspondingauthor{Lou Strolger}
\email{strolger@stsci.edu}

\author[00000-0002-7756-4440]{Louis-Gregory Strolger}
\affiliation{Space Telescope Science Institute, Baltimore, MD 21218, USA}
\affiliation{Johns Hopkins University, Baltimore, MD 21218, USA}
\affiliation{Morgan State University, Baltimore, MD 21251, USA}

\author[0000-0002-6042-1486]{Jamila Pegues}\altaffiliation{STScI Postdoctoral Fellow}
\affiliation{Space Telescope Science Institute, Baltimore, MD 21218, USA}

\author[0000-0001-8031-8606]{Tegan King}
\affiliation{National Center for Atmospheric Research, Boulder, CO 80305, USA}

\author[0000-0001-8936-4545]{Nathan Miles}
\affiliation{Department of Earth, Planetary, and Space Sciences, University of California, Los Angeles, Los Angeles, CA 90095, USA} 

\author{Michelle Ramsahoye}
\affiliation{Department of Mathematics and Statistics, University of Maryland, Baltimore County, Baltimore, MD 21250}

\author{Keith Ceruti II}
\affiliation{Baltimore Polytechnic Institute, Baltimore, MD 21209, USA }

\author{Brett Blacker}
\affiliation{Space Telescope Science Institute, Baltimore, MD 21218, USA}

\author[0000-0003-0531-8547]{I.~Neill Reid}
\affiliation{Space Telescope Science Institute, Baltimore, MD 21218, USA}

\begin{abstract}

With the start of a new Great Observatories era, there is renewed concern that the demand for these forefront facilities, through proposal pressure, will exceed conventional peer-review management's capacity for ensuring an unbiased and efficient selection. There is need for new methods, strategies, and tools to facilitate those reviews. Here, we describe PACMan2, an updated tool for proposal review management that utilizes machine learning models and techniques to topically categorize proposals and reviewers, to match proposals to reviewers, and to facilitate proposal assignments, mitigating some conflicts of interest. We find that the classifier has cross-validation accuracy of $80.0\pm2.2\%$ on proposals for time on the {\it Hubble Space Telescope} and the {\it James Webb Space Telescope}.
\end{abstract}

\section{Introduction} \label{sec:intro}
 PACMan~\citep{Strolger:2017tt} was initially conceived as a means to reduce subjective bias and improve efficiency in the recruitment of appropriate reviewers for proposals to the \textit{Hubble} and \textit{James Webb} space telescopes (\hubble{} and \webb{}, respectively). The Space Telescope Science Institute (STScI) conducts its peer reviews of proposals for telescope time in a process generally similar to those used worldwide, for both space and ground-based observatories. Proposals are solicited through Calls for Proposals (CfPs), released annually as Cycles for each mission. Those proposals are evaluated on their scientific merit by a Time Allocation Committee (TAC), a panel of peers recruited for their expertise in broad topical science areas. On average, over $1000$ proposals are received in response to each CfP, of which only the top $\sim20\%$ are selected. The Science Policies Group (SPG) at STScI arranges these reviews, recruiting a few hundred experts from the international scientific community in a six-month process that culminates in a set of ranked recommendations to the STScI Director.
   
As anticipated, \webb{} has essentially doubled the number of proposals received each year, greatly increasing the pressure in arranging thorough and unbiased reviews. There is just a finite number of qualified and available reviewers in the community, and that pool is further limited by our institutional awareness of who they are and what they do. Moreover, the global COVID pandemic has generally further reduced reviewer availability worldwide.  In anticipation of a shortfall, with \cite{Strolger:2017tt} we began to address how machine-learning tools would help streamline the review process, and to do so without increasing the potential for biases typically encountered in pressured reviews. The work was focused on automating and improving reviewer matching, and thereby expanding our reviewer pools, but still working within our peer-review framework. We also hoped not to impart subjectivity worse than as seen for typical expert reviews~\citep{Patat:2018ag}. It does not consider changes to the way in with the peer reviews are done, e.g., like distributed peer-reviews~\citep{Kerzendorf:2020um} similar to those employed at Gemini and ESO for some parts of their biannual reviews, as there are  shared concerns on strategic manipulation~\citep{Naghizadeh:2013fg}. 

Here we present the latest incarnation of the PACMan (version 2) tool-kit\cite{King:2023aa}, providing a more direct proposal-to-reviewer matching, and supporting tools to help manage the reviews. PACMan2 has a set of operations to assist in the creation of panel reviews and proposal assignments. In this paper we document its core functions: topically classifying proposals and reviewers, identifying reviewers whom are best suited to reviewing specific proposal, determining close collaborators and potential conflicts of interest, and identifying resubmissions. 

As with the previous version, PACMan2 uses Naive Bayesian techniques. Open-sourced python-based machine-learning toolkits such as \texttt{scikit-learn}~\citep{scikit-learn} have become more prevalent in the community, providing robust tools for doing all sorts of predictive analysis and classification. These tools are known to work particularly well when combined with Natural Language Processing techniques such as word-count vectorization~\citep{Kerzendorf:2019zc}, for which toolkits have also become more widespread~\citep[e.g., \texttt{spaCy}, ][]{spacy}. PACMan now utilizes the \texttt{scikit-learn} naive Bayes algorithm for multinomially distributed data, \texttt{MultinomialNB}, and takes advantage of the \texttt{spaCy} tools for lemmatization, tokenization, and word vectorization, with stop words.

\section{Improved Proposal and Reviewer Categorization}\label{sec:categorization}
PACMan's classifications are made with a multinomial naive Bayesian classifier, in a manner similar to \cite{Strolger:2017tt} and \cite{Kerzendorf:2019zc}, using word vectors with Term Frequency-Inverse Document Frequency (TF-IDF) weighting statistics.  In brief, the Bayes' method determines the probability of a corpus (or proposal) belonging to a pool or topical category, from the joint probability of words used in that corpus belonging to those same categories from frequency of use in training sets. TF-IDF down-weights words commonly used across multiple categories. 

We classify proposals into broad pools based on science topics, correlating with the topical panels of the 2022 TAC. Those are Solar System Astronomy (abbreviated as Sol. in some areas of this text), Exoplanets and Exoplanet Formation (or Exopl.), Stellar Physics and Stellar Types (Stars), Stellar Populations and the Interstellar Medium (StPops), the Intergalactic Medium and Circumgalactic Medium (IGM), Supermassive Black Holes and Active Galaxies (AGN), Galaxies (Gal.), and Large Scale Structure of the Universe (LSS). These terms were chosen largely due to their adherence with the Unified Astronomy Thesaurus~\citep[UAT, ][]{Frey:2018ux}, which could allow for a simple, hierarchical connection of related sub-topics, or from one topical categorization to the next. This is put into practice in the recruitment of panelists who select from options of Scientific Categories and Keywords, similar to Table 3 of \cite{Frey:2018ux}, to narrow in (at a finer level) on their areas of expertise, to more directly match by hand to a similar keywords and categorization selections provided by proposers. We have only modestly tested categorizing with PACMan at such a refined level, the results of which, as shown in Figure 5 of \cite{Strolger:2017tt}, tends to show a much higher misclassification (or failure) rate that does not significantly improve with the size of the training pool. 

\begin{figure}[t]
\begin{center}
\includegraphics[width=0.9\textwidth]{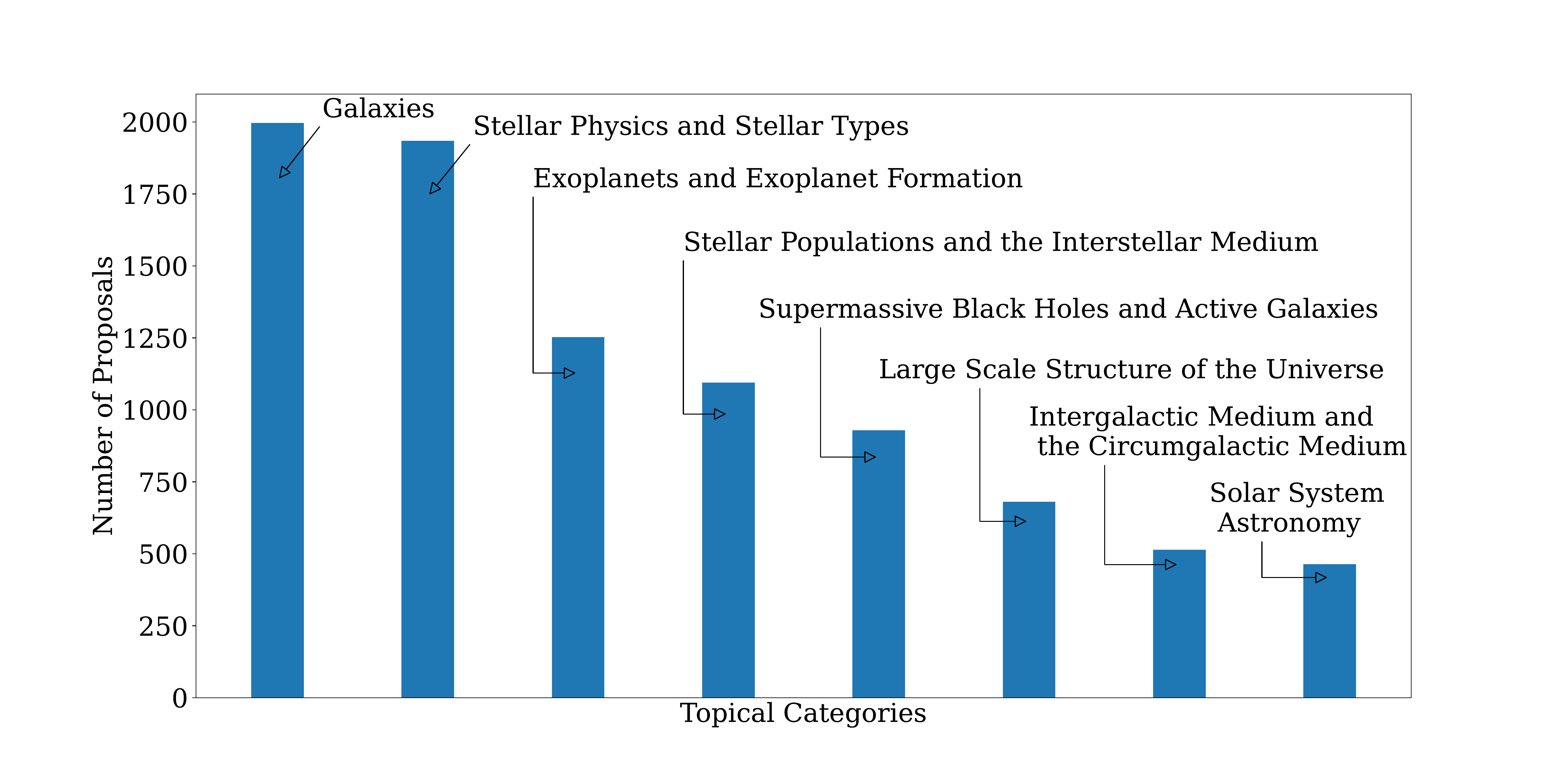}
\caption{Topical classifications of the proposals in the training set. Proposals were initially categorized by the proposers according to the topical panels by which they desired the proposals to be reviewed. Proposals prior to Cycle 30 required minor re-categorization (by the authors of this manuscript) to adhere to the categories used in Cycle 30.~\label{fig:handclass}}
\end{center}
\end{figure}

Our classifier model is built on training from 8868 proposals received from the community in \hubble{} Cycles 23 through 30, for the regular CfP and for mid-cycle opportunities, in the years 2015 to 2022. The proposer-selected science categories of those proposals are shown in Figure~\ref{fig:handclass}, where proposals from earlier cycles were re-mapped to fit the science topics above.

\subsection{Cross Validation}

\begin{figure}[t]
\begin{center}
\includegraphics[width=0.7\textwidth]{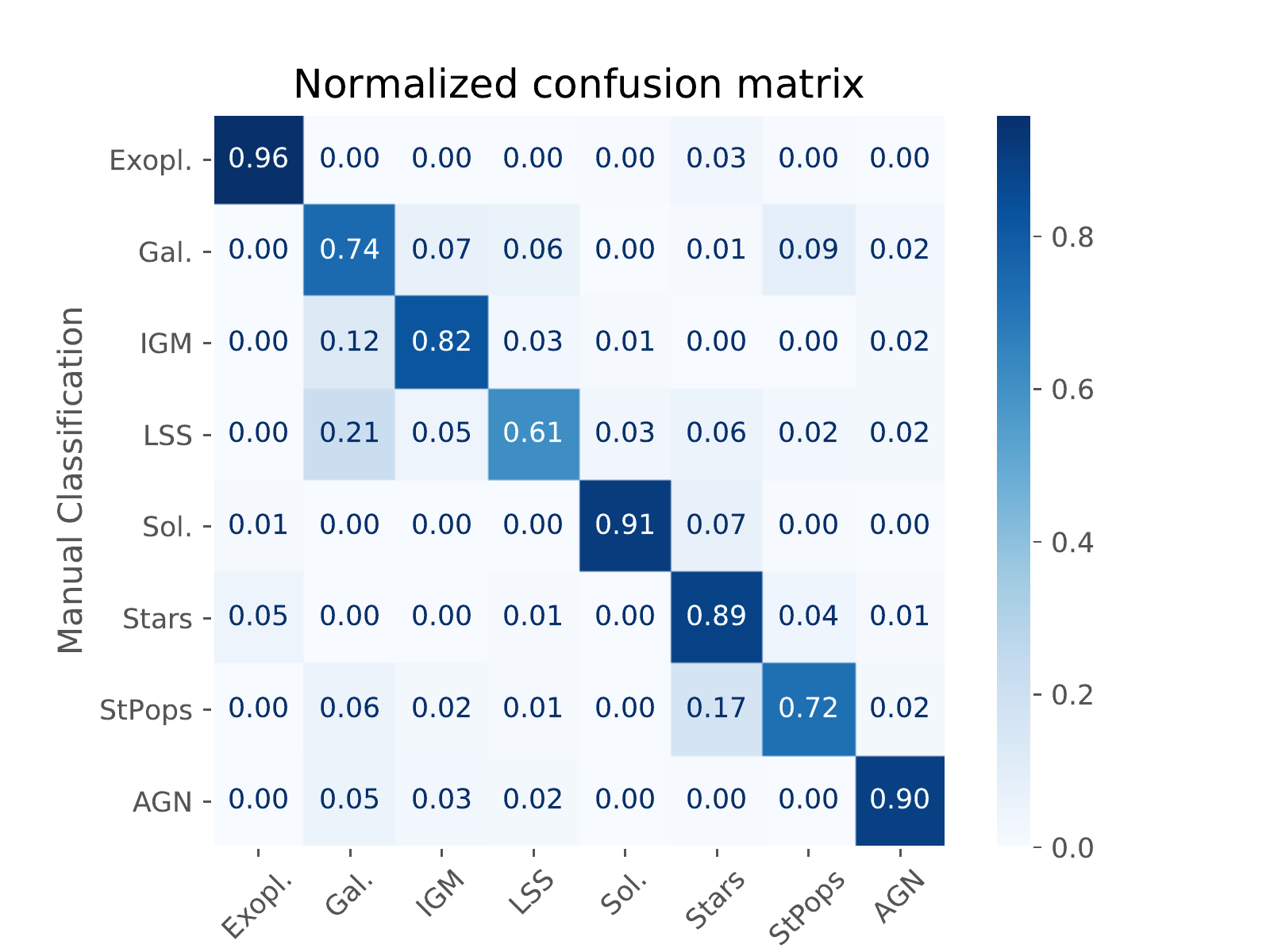}
\caption{\footnotesize Normalized confusion matrix of the classification model, determined from cross-validation testing. Some topical categories, e.g., Large Scale Structure of the Universe, could be confused with closely related topics, such as Galaxies.\label{fig:conf_mat}}
\end{center}
\end{figure}

\begin{figure}[t]
\begin{center}
\includegraphics[width=0.6\textwidth]{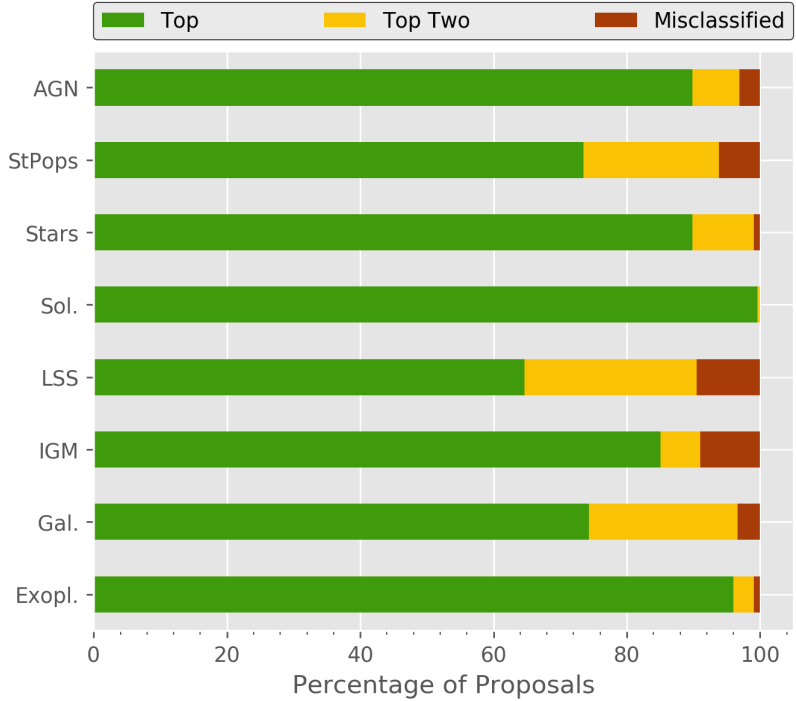}
\caption{\footnotesize Similar to Figure 3 of \cite{Strolger:2017tt}, a comparison (or precision of) the PACMan categorization to the manual sorting of proposals from \hubble{} Cycles 23$-$30. The percentage of proposals that are similarly categorized by both are given in green. The yellow indicates how often PACMan's second choice is the same as the `by hand' classifications in that category. The red indicates the frequency of `misclassifications', when the manual classification is not within PACMan's top two choices.\label{fig:barh}}
\end{center}
\end{figure}

We tested our classifier model via a cross-validation method, using a \textit{leave-one-out approach} where we specify 7 stratified K-folds of testing data, where in machine-learning parlance, $k$ indicates the number of samples for each test (i.e., $k=7$). Here, the data is randomly split into a training set (of 85.7\% of the proposals) with stratified equal numbers of training proposals in each of the 8 categories, from which a temporary model is then derived. That model is then evaluated and scored on the remaining 14.3\% of the sample (i.e., the test set). The process is repeated 7 times, making new stratified splits of training and testing along a 6:1 ratio, discarding the temporary model while retaining the score for the aggregate summary. The cross-validation accuracy of the classifier from this testing is $80.0\pm2.2\%$, while the precision and f1-measure (the harmonic mean of the precision, the fraction of categorized corpora that were accurate, and recall, the fraction of relevant corpora were accurately categorized) are $78.8\pm2.7\%$ and $78.2\pm2.4\%$, respectively. The normalized confusion matrix, shown in Figure~\ref{fig:conf_mat}, emphasizes where mis-categorizations occur, in closely related topical categories. For example, Large Scale Structure of the Universe shows the largest misclassification rate (39\%), with the most frequent confusion being with proposals in the Galaxies category (21\% of the time). We attempt to capture the efficacy of the `top two' classifications, in a method similar to Figure 3 of ~\cite{Strolger:2017tt}, in Figure~\ref{fig:barh}. Here, the frequency in which PACMan matches the manual classifications are shown in green, and how often PACMan's second choice matches the manual classifications is shown in yellow. The `misclassifications', when the manual classification is not within PACMan's top two choices, are shown in red.

As before, it is not surprising that PACMan does best in categories in which there are word tokens which unambiguously belong to said categories, such as planet names in Solar System Astronomy, or ``coronagraphy" in Exoplanets and Exoplanet Formation. Topics which give PACMan the most difficulty, i.e., Stellar Populations, Galaxies, and Large Scale Structure of the Universe, have also been areas of great ambiguity to proposers.  Nonetheless, PACMan has an average `top category' accuracy of 84\%, and a `top 2 categories' average accuracy of 96\%. PACMan does no worse than about a 10\% classification error in any of the 8 topical categories presented. As a note, while  \texttt{scikit-learn} provides a naive Bayesian method for imbalanced training sets, which may better handle our data as implied by Figure~\ref{fig:handclass}, our testing of the Complement Naive Bayes classification shows very little (but slightly worse) difference in overall classification performance, with a cross-validation accuracy of $76.9\pm2.7\%$, and little change in the confusion matrix.

\subsection{Characterizing Reviewer Domain Knowledge}\label{sec:domain}
Until relatively recently, recruiting reviewers for the TAC has relied solely on the SPG's familiarity with the field, and knowledge of the researchers who are active within it. A few years ago the Proposal/Person (or ProPer) Application\footnote{ProPer can be found on \url{https://proper.stsci.edu/proper/authentication/}.} and its associated database was developed to maintain contact information for people who have applied for time with \hubble{} or \webb{}, or those associated with the STScI-organized service committees, largely for \hubble{} and now for \webb{}.  Since then, the interface has been used internally to keep a record of self-declared areas of scientific expertise. ProPer has very much facilitated the search for panelists to serve on the TAC topical panels, and through the addition of UAT terms as keywords, matching proposals to reviewers has also been made easier (more on that in Section~\ref{sec:matching}).  Yet, this extensive record only exists for those who have applied for time with these observatories, or in one way or another, connected to STScI service committees. It does not reach to, for e.g., primarily ground-based astronomers, or astronomers focusing on other wavelengths, or in many cases, theorists, instrumentalists, or other non-observers. These investigators are likely equally knowledgeable about the science to be evaluated, without being experts on the use of these facilities. We wish to apply our trained model to categorize potential reviewers under the same pools as proposals, with a goal of facilitating and expanding potential reviewer pools for each of the appropriate topical review panels.

\begin{figure}[t]
\begin{center}
\includegraphics[width=0.55\textwidth]{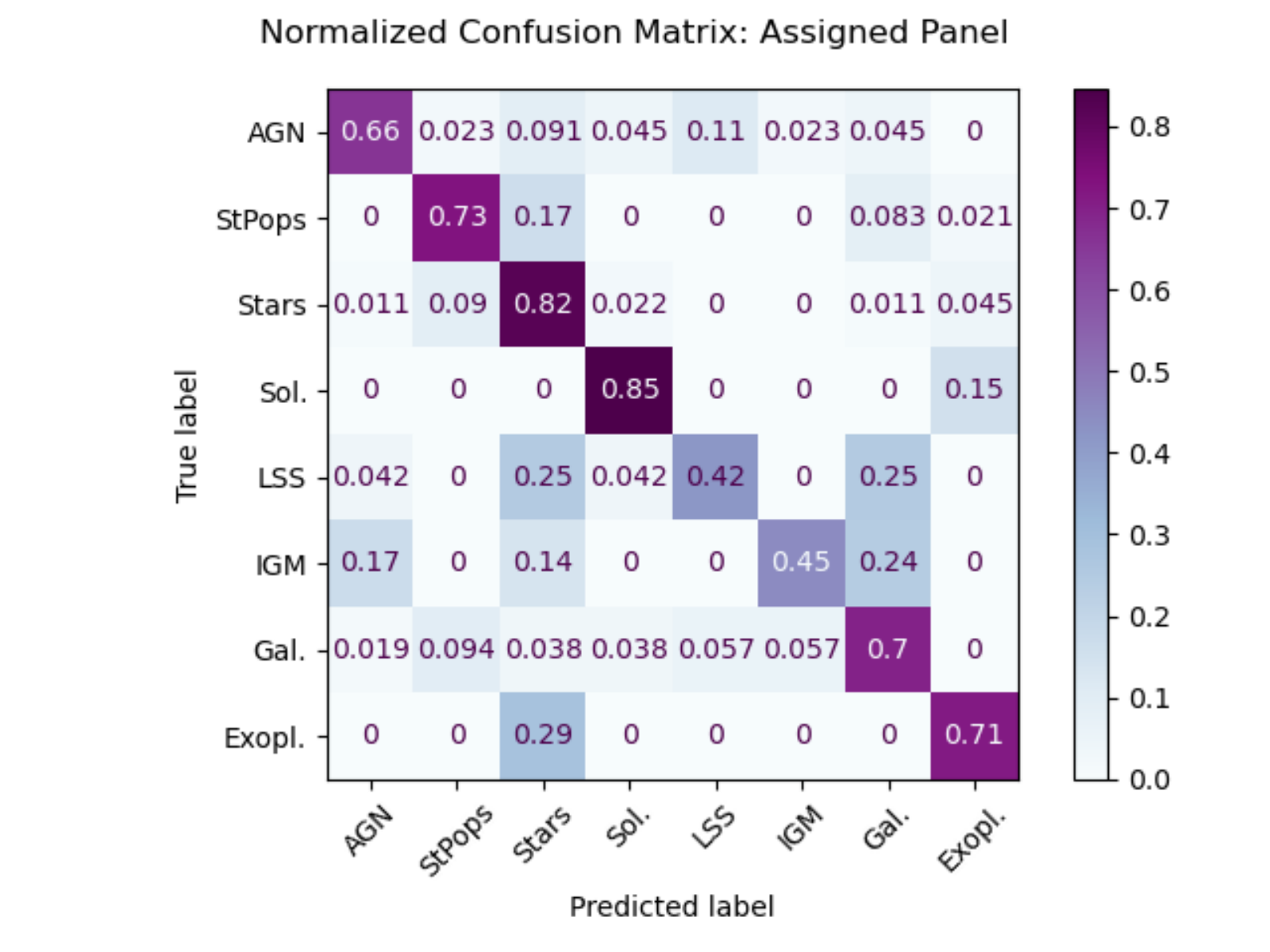}
\includegraphics[width=0.43\textwidth]{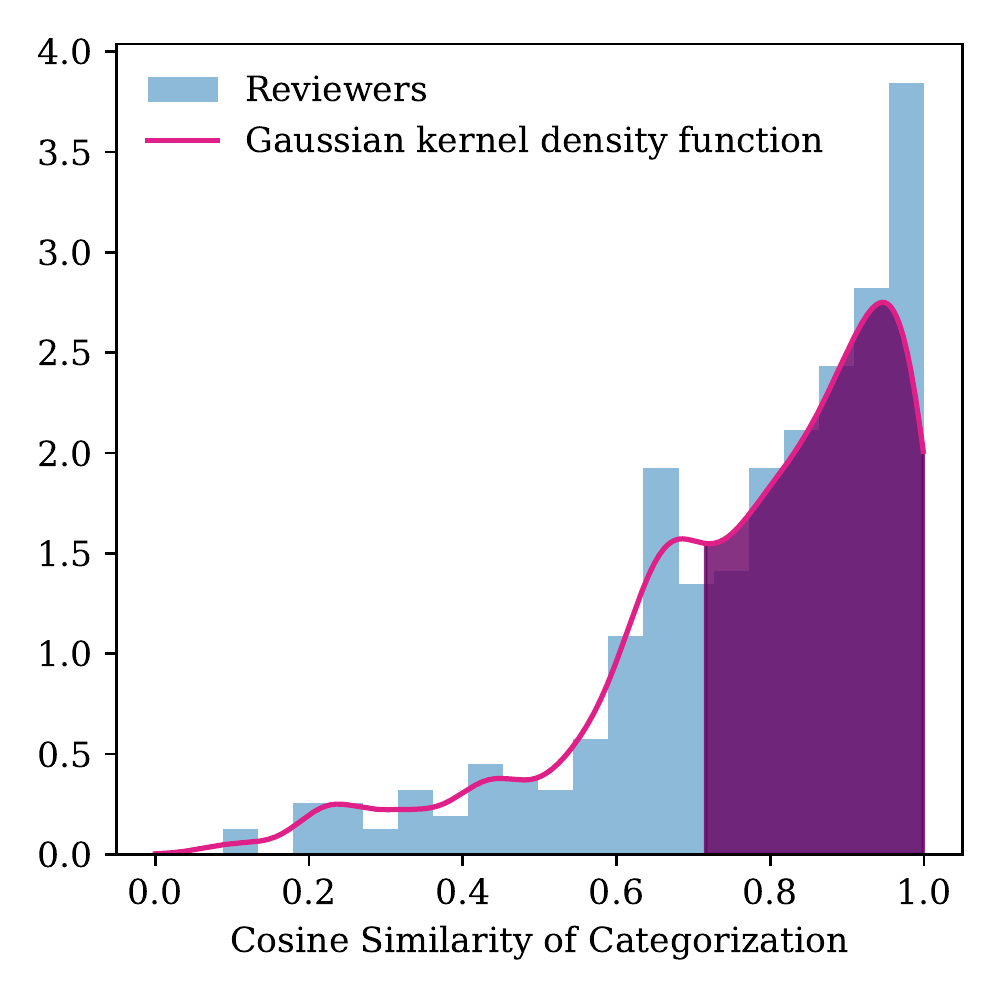}
\caption{\footnotesize Comparisons of reviewer categorizations via PACMan (predicted label) to those provided by the reviewers themselves (true label). \textit{Left:}~the normalized confusion matrix between reviewer-supplied categories of expertise and PACMan category vectors on reviewers. \textit{Right:}~normalized distribution of cosine similarity between category vectors. The dark purple region marks the top $68^{th}$ percentile of the gaussian kernel distribution (purple line).\label{fig:user_cats}} 
\end{center}
\end{figure}

Assuming a record of domain knowledge exists in an author's bibliography, we create evaluation corpora from Astrophysics Data System (ADS)\footnote{\url{https://ui.adsabs.harvard.edu}.} queries for the last 10 years worth of refereed paper abstracts, and in lieu of proposals, from a list of proposed referees. These authors are identified in queries to ADS (made via the API\footnote{\url{https://ads.readthedocs.io.}}) by first name and last name. It would be ideal to use a persistent identifier (e.g., ORCID\footnote{\url{https://orcid.org}. The public API can be found on \url{https://info.orcid.org/documentation/features/public-api/}. } iDs) that distinguishes authors with the same or similar names, and unambiguously identifies those who may have changed their names over their careers. But unfortunately, from our experience, the record of ORCID iDs has been incomplete for many authors in our field, and therefore an unreliable query to ADS. The occasional ambiguous query from names alone results in more records attributed to a given name than are warranted, and a broader characterization (and more frequent proposal matching, see Section~\ref{sec:matching} below) than would be expected. At the moment, there is little to guard against this, and therefore these results (as with all PACMan results) should be used as recommendations to guide to reviewer-panel organizers and managers. We remain hopeful that the usage of such identifiers expands greatly, and envision incorporating the ORCID public API with our code in a future release.

PACMan interprets these corpora, applying the same lemmatization, tokenization, and word vectorization as is done with proposals, and the same trained classification. As an illustration of the effectiveness of this method, Figure~\ref{fig:user_cats} compares reviewer categorizations from PACMan to those supplied by the reviewers themselves for \hubble{} Cycle 30. Here, the internal and external reviewers provide their expertise in ProPer through selections of sub-category keywords.

As a test of the PACMan-assigned domain knowledge, against user-supplied information, we make use of the fact that it is often the case that reviewers have expressed at least some expertise in a wide-range of keywords that map-back to more than one topical category. Moreover, PACMan categorizations are actually probabilities of expertise in each topical category. Both tidbits of information can be used and compared, converting the frequency of times certain categories are selected from mapped-back keywords into a mock category vector ($\vec{v_1}$), and using the PACMan probabilities as a separate mock category vector ($\vec{v_2}$), we can compute the cosine similarity of these two mock category vectors as:
\begin{equation}
	S_C(\vec{v_1},\vec{v_2})=\frac{\vec{v_1} \cdot \vec{v_2}}{\|\vec{v_1}\| \|\vec{v_2}\|}.\label{eqn:cosine}
\end{equation}
\noindent The right panel of  Figure~\ref{fig:user_cats} shows the cosine similarity ($S_C$) distribution, which indicate that the PACMan categorizations agree, for the most part, with the reviewer-supplied categorizations. In finer detail, these 8-component $S_C$'s are largely very high (where $S_C\equiv1$ is a perfect match), with the highest 68\% at $S_C>0.7$. These mock 8-component vectors were only constructed for this test of domain knowledge, and are not used elsewhere in the PACMan codebase. Instead, the multi-component word vectors, as further described in Section~\ref{sec:matching}, are used and are naturally more complicated, and as such result significantly lower (i.e., $S_C\ll1$) values.

It's not always straight-forward for an individual to characterize their expertise in limited topical categories. Further, there are known biases in how individuals characterize the scope of their work. PACMan potentially captures a better, less-biased characterization of that expertise from the publications these individuals author and co-authors. The interesting collateral outcome of these model-based characterizations may be the result of a wider range of expertise, for a given reviewer, than said reviewer may have provided themselves. 

\section{Reviewer-to-Proposal Matching}\label{sec:matching}
The categorization tool provides a base method for identifying which topical panel proposals should be sent to, and who would be adroit to serve on those review panels. The next step is direct proposal-to-reviewer assignments. The \hubble{} reviews previously made use of the user-defined categories and keywords, and assignments were primarily made based on the number of proposal-reviewer overlapping keywords. Other considerations, such as load balance and reducing the conflicts of interests (discussed in Section~\ref{sec:conflicts}) would further solidify those assignments. However, experience has shown that user-provided categorizations and keywords can be misused, misconstrued, or underutilized, leading to inaccuracies in those assignments.

There are well-established methods for pairwise comparisons of corpora, determining likenesses (or affinities), or differences. One simple evaluator is the cosine similarity ($S_C$), as described in Equation~\ref{eqn:cosine}. Forgoing the categorizations described in Section~\ref{sec:categorization}, and the mock 8-component vectors from the Section~\ref{sec:domain}, this method directly compares the TF-IDF word-frequency multi-component vectors to each other to  determine which corpora are best aligned with a given reviewer's ADS bibliography (as described in Section~\ref{sec:domain}). While $S_C$ scores closest to a value of 1 represent the most ideal matches, it would be nearly impossible for a proposal to use \textit{exactly} the same words, or in the same frequency, as those used in the manuscripts of an author's bibliography. It begs the question of what range of $S_C$ values which represent scores for `experts', or those best qualified to review a given proposal?

\begin{figure}[t]
\begin{center}
\includegraphics[width=\textwidth]{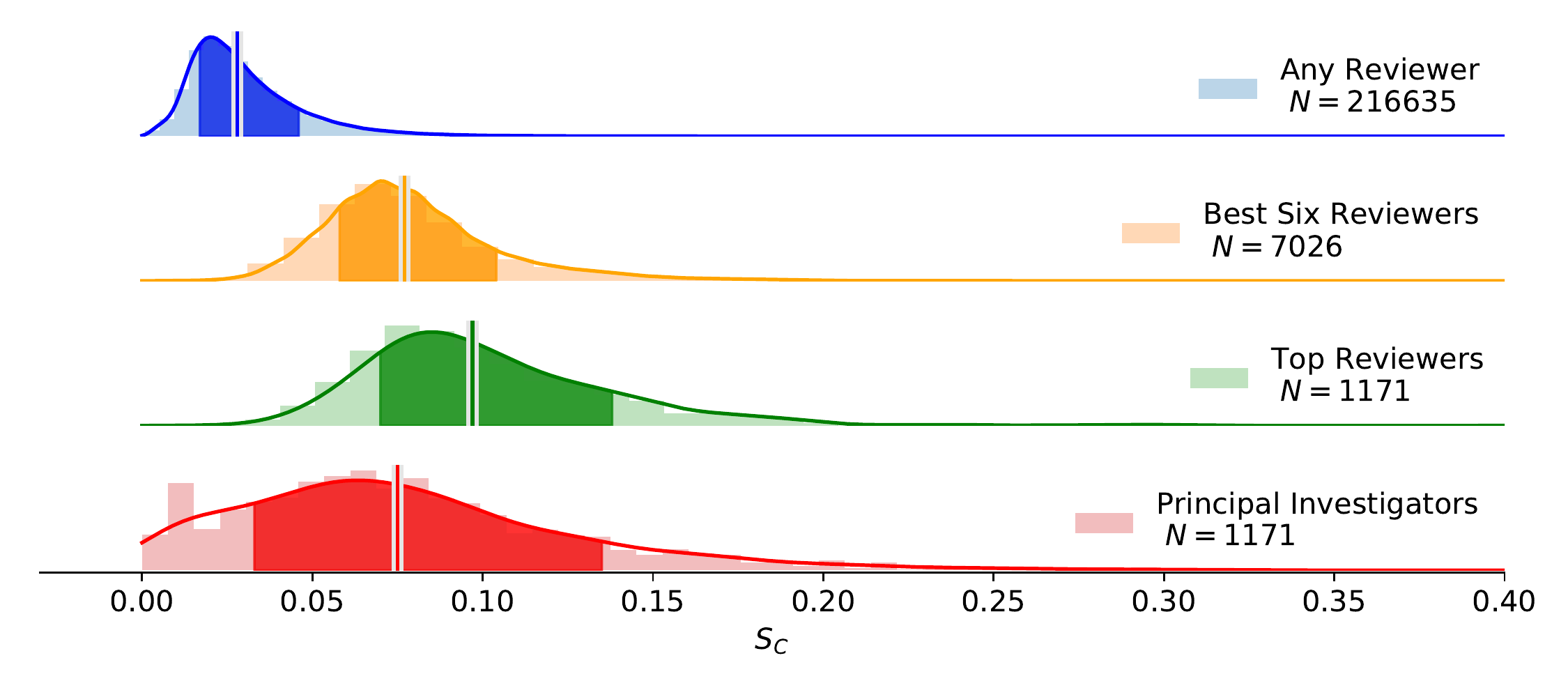}
\caption{Distributions of proposal-to-reviewer cosine similarity ($S_C$) scores are shown, along with their gaussian kernel distributions, for the subgroups of reviewers selected by expertise shown in the legend and in Table~\ref{tab:1}. Vertical lines indicate the median $S_C$ values, and shaded regions show the 68\% confidence intervals around those medians. The number of within those subgroups are also indicated in the legend.\label{fig:1}} 
\end{center}
\end{figure}

To test this we sought to compare the list of potential reviewers generated through ProPer for the \webb{} Cycle 1 TAC review, against the proposals received for that CfP deadline in 2020. We constructed a few tests of this list of reviewers against the proposals, the first evaluating all 185 potential reviewers against all 1171 proposals, totaling 216,635 scores, to create a baseline of $S_C$ score distribution (see Figure~\ref{fig:1}). As one would expect, some subset of reviewers have the highest scores on a few proposals, with the top 15\% in the range $S_C>0.046$. But most comparisons would be average to poor, with $\bar{S}_C=0.03^{+0.02}_{-0.01}$ and $\rm{Mo}(S_C)=0.02$. To better isolate the range best scores, we looked more closely at just those top-scoring reviewers for each proposal, then the highest $S_C$, and then the top six reviewers assigned to each proposal\footnote{Six was somewhat arbitrarily chosen. Proposals are generally assigned to three readers, which was doubled to guard against conflicts of interest which could disqualify some readers.}. The distribution of those $S_C$'s for the top-six reviewers was clearly distinguished from the baseline, with $\bar{S}_{(C,\uparrow 6)}=0.08^{+0.03}_{-0.02}$ and $\rm{Mo}(S_{C,\uparrow 6})=0.07$. And the best reviewers tended to even higher scores, with $\bar{S}_{(C,\,{\rm Best})}=0.10^{+0.04}_{-0.03}$ and $\rm{Mo}(S_{C,\,{\rm Best}})=0.08$. These results are also shown in Table~\ref{tab:1}. From the distributions, it seems $S_C\ge 0.06$ generally denotes a range in which a reviewer is sufficiently expert enough to review a given proposal, and that assignments of $S_C<0.05$ should be done with some caution.

\begin{table}[ht]
   \centering
   \begin{tabular}{@{} lcr @{}} 
      \toprule
      	 & $\bar{S}_C$ & $\rm{Mo}(S_C)$ \\
	 \hline
	Any Reviewer & $0.03^{+0.02}_{-0.01}$ & 0.02\\
	Best Six Reviewers ($\uparrow 6$) & $0.08^{+0.03}_{-0.02}$ & 0.07\\
	Top Reviewers (Best) & $0.10^{+0.04}_{-0.03}$ & 0.08\\
	PIs as Reviewers (PIs) & $0.08^{+0.06}_{-0.04}$ & 0.05\\
      \hline
   \end{tabular}
   \caption{Medians, 68\%-confidence regions, and modes for the subgroups of reviewers selected by expertise, in the evaluation of range of validity of proposal-to-reviewer cosine similarities ($S_C$).  }
   \label{tab:1}
\end{table}

It is not always the case that the proposers themselves are the best suited to evaluate the subject matter of their proposals. Figure~\ref{fig:1} also shows the distributions of $S_C$ scores for the proposal principal investigators (PIs) against their own proposals, showing median expertise on par with the best six reviewers, but with a wider distribution, with $\bar{S}_{(C, {\rm PIs})}=0.08^{+0.06}_{-0.04}$. So, while it encompasses the range of some of the top-most reviewers, it also covers some of the range of the least expert of reviewers. 

This exemplifies some issues in how $S_C$ is determined, and perhaps serves as a caution for using these values wholesale without due consideration. It is very possible for any author to have expertise in several areas, and to develop expertise in emergent areas. It is this latter case that PACMan would be particularly poor at, determining their overall expertise on a topic for which they haven't written much about. PACMan can, however, respond with who amongst the pool of names given is best to review a given proposal, which would be a proper interpretation of its results.

\section{Identifying Close-collaborators as Potential Conflicts of Interest}\label{sec:conflicts}
The categorization models and $S_C$ tools described above provide a great starting point for recruiting panels and making reader assignments. However, such assignments should also be weighed against other considerations, like assignment load, and more importantly, potential for conflicts of interest. Generally, conflicts of interest arise when a reviewer's interest, in direct or tractable resource (or financial) gain, is at odds with their ability to provide an unbiased review of a given proposal. This definition can be expanded to cover even the perception of conflict of interest, to an outside observer, even though the actual gain to the reviewer may be minimal. 

The SPG at STScI has taken great strides to reduce such biases, implementing a dual-anonymous peer review\footnote{See \url{https://outerspace.stsci.edu/display/APRWG}.} in which the identities of the proposers are not directly known to the readers as they do their assessment of scientific merit. On the one hand, it would seem that such a policy addressing identity bias would have the collateral benefit of essentially eliminating conflicts of interest, \textit{ab initio}-- if the reviewer does not know the identities of the proposers, how would they know if they stand to gain from its award or rejection? But on the other hand, it is important to note that the goal of those dual-anonymous policies are not to completely erase all identity of the proposers, rather to obscure them enough so that those identities are not a focus of discussion during the review of scientific merit. There may yet be room for conflicts of interest to arise. The unfortunate side effect of the process is that readers cannot simply review the authorship and directly know if they have a conflict of interest to declare. The onus of determining these conflicts defaults back to the SPG and requires tools to help facilitate. 

The queries to the ADS interface used in panel assignments and reviewer assignments also provide a means for identifying `close-collaborators', at least in published works. From these queries, PACMan provides lists of close collaborators with a given reader on three escalating levels: any co-authors, co-authors on 3 or more papers, and co-author as a primary author (any of the first three names) in the past 3 years. These criteria provide a range of minor, moderate, and major conflicts of interests, respectively, which are appropriately weighed with PACMan-recommended panel and reviewer assignments by the SPG in finalizing actual assignments.

\section{Determining Resubmissions}
It is often very useful to know what proposals are re-worked resubmissions from previous cycles, and without an encyclopedic knowledge of every proposal ever received, those can be very hard to identify. PACMan provides a tool to determine what proposals in past submissions have close similarity to those submitted for the current cycle. Those proposal duplications are determined via hashed winnowed $k$-gram similarity, using the method of \cite{Citron:2015uk}, developed for \href{ar$\chi$iv.org}{arXiv.org}. Using $k$-grams, or sentences of greater than $k$-words length, the method essentially creates a fingerprint of each proposal, by collecting the set of $k$-grams within each. A proposal of $n$ words has $n-(k+1)$ k-grams. Using \texttt{Fingerprint}\footnote{\texttt{Fingerprint} codebase: \url{https://github.com/kailashbuki/fingerprint.} under an MIT license.}, each $k$-gram is hashed into a unique long integer identifier by using the remainder of dividing the $k$-gram by \texttt{sys.maxsize}\footnote{The maximum size object containers (lists, strings, etc.) can have in python, which for 64-bit machines is $2^{63}-1=9,223,372,036,854,775,807$. The method is discussed in great detail in the \texttt{Footprint} GitHub repository.}, with the string of hashes forming a corpus's ``fingerprint'', easily indexed in a computer's RAM. One can further facilitate calculations by taking advantage of the fact that $k$-grams have overlap, and therefore redundant information that unnecessarily increases the size of these fingerprints. The hashes are thus winnowed by a factor of $t>k$, where each proposal is composed of $n-t+1$ windows, each with $t-k+1$ winnowed $k$-grams.  Adopting the lowest hash in rolling windows of length $t$ further compresses the size of corpora fingerprints. Based on \cite{Citron:2015uk} recommendations, we use $k=7$ and $t=12$, compressing fingerprints of winnowed 7-grams (w7g's) effectively by factors of 3 to 4.

\begin{figure}[t]
\begin{center}
\includegraphics[width=\textwidth]{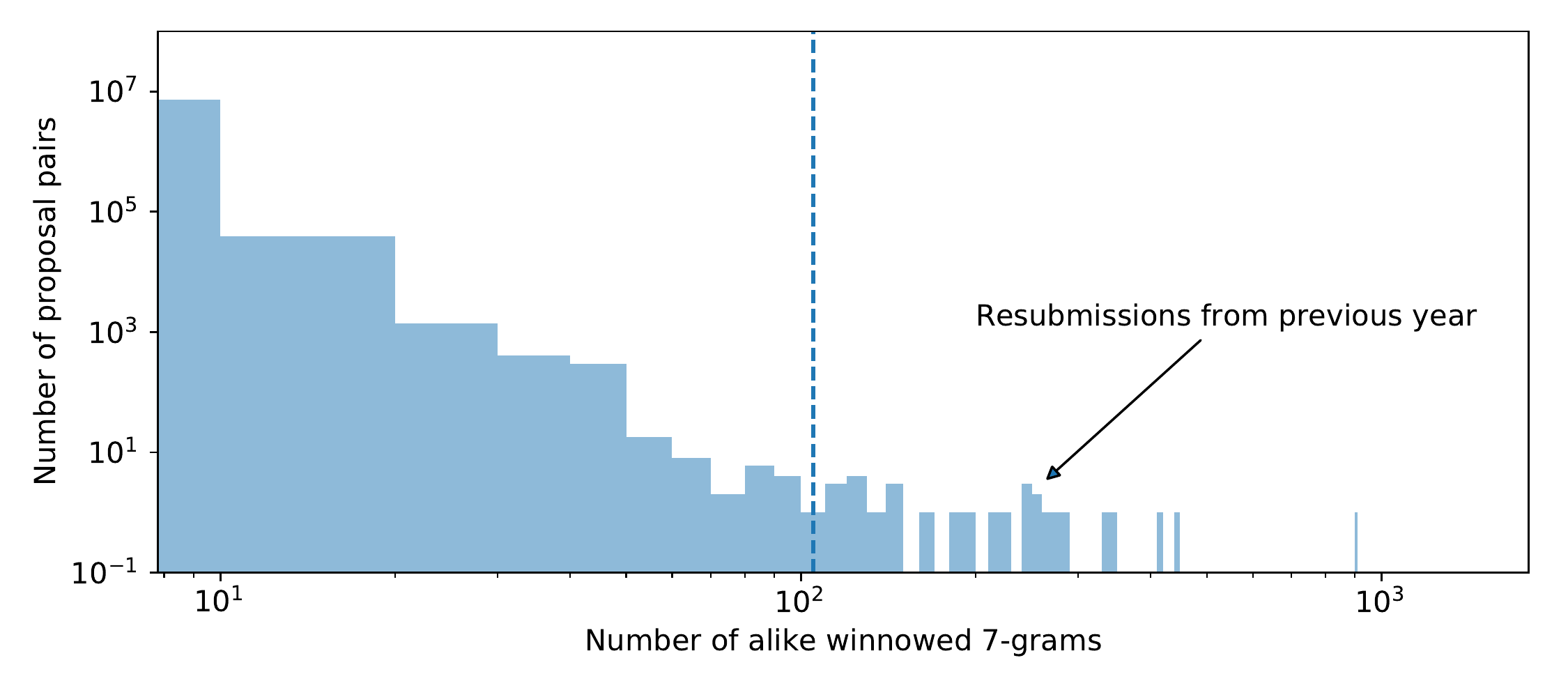}
\caption{The number of proposal pairs with $N$ alike winnowed 7-grams (w7g) from Cycle 29, compared against Cycles 23 through 29. From manual reviews, proposal pairs that share more than one-hundred w7g in common are essentially resubmissions.\label{fig:2}}
\end{center}
\end{figure}

\begin{figure}[t]
\begin{center}
\includegraphics[width=\textwidth]{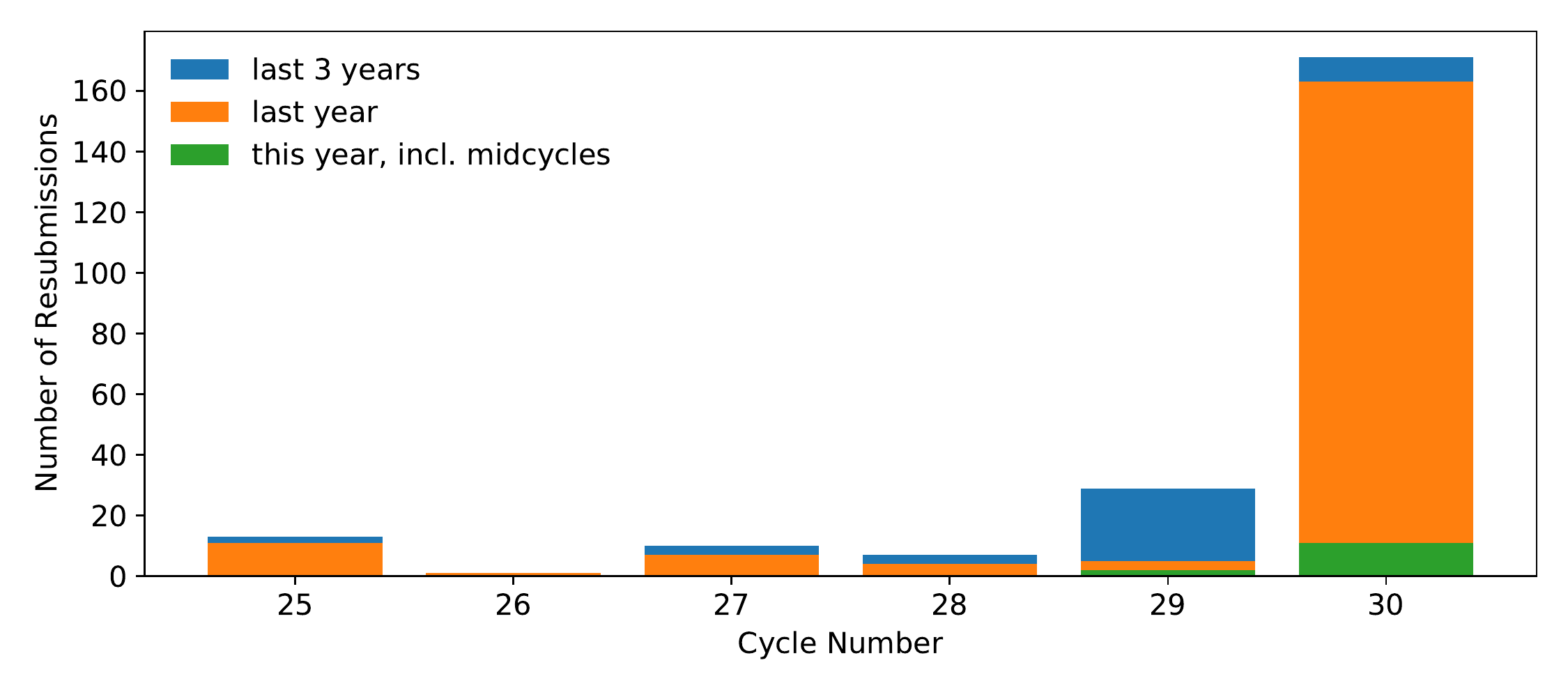}
\caption{The number of suspected resubmissions (w7g $> 100$)  in the last six cycles, shown as a stack histogram from within the same year (from mid-cycle proposals, in green), then including resubmissions from the last cycle (in orange), then from $\le3$ years (in blue).\label{fig:3}}
\end{center}
\end{figure}

Figure~\ref{fig:2} shows the distribution of w7g values for the 1130 proposals received in Cycle 29, in comparison to the 7805 proposals in Cycles 23 through 29. Similar distributions were drawn for each Cycle, 23 through 30, and reviewed manually, from which it was determined that w7g $\gtrsim100$ were highly likely resubmissions. From these manual reviews, it was also noted that the number of resubmissions appeared to be growing. Adopting w7g $>100$ as the cutoff, we show (in Figure~\ref{fig:3}) that the number of suspected resubmissions has gone up in the last 5 years, most notably in Cycle 30. Most of the resubmissions tend to be from just the previous cycle ($\approx80\%$), with most of the remainder coming from within the last three cycles ($\approx14\%$). The subsequent remainder of $\approx 6\%$ are resubmissions of mid-cycle proposals.

\section{Summary}
PACMan provide a useful toolkit for proposal and reviewer management in the STScI peer-review system, and similarly should have broad applicability to other peer-review management groups. The PACMan version 2 codebase is available on zenodo~\citep{King:2023aa}, and maintained on \texttt{github}, \url{https://github.com/spacetelescope/PACMan}.

\acknowledgements{We thank our anonymous referee for important improvements to this manuscript. We also thank Alessandra Alossi, Claus Leitherer, Christine Chen, Kathrine Alatalo, Molly Peeples, and Laura Watkins for their discussions in preparing and implementing PACMan.  MR thanks the STScI Space Astronomy Summer Program. KC thanks The Baltimore Polytechnic Ingenuity Project for enabling their participation in this work. This research has made use of NASA's Astrophysics Data System Bibliographic Services, in particular, the API for the Astrophysics Data System.}

\vskip6pt
{\large\it Author contributions:} We use the CRT standard (\url{https://casrai.org/credit/}) for reporting author contributions. Conceptualization: L.G.S. Data curation: L.G.S., B.B., I.N.R. Formal analysis: L.G.S, J.P., K.C. Investigation: L.G.S., J.P., T.K., N.M., M.R., K.C., B.B., I.N.R. Methodology: L.G.S. Software: T.K., N.M., M.R. Supervision: L.G.S. Validation: L.G.S. Writing$-$original draft: L.G.S. Writing$-$review \& editing: L.G.S., J.P., T.K., N.M., M.R., K.C., B.B., I.N.R.

\facilities{\textit{Hubble Space Telescope}, \textit{James Webb Space Telescope}}
\software{\texttt{Astropy} \citep{Astropy:2013},
\texttt{Matplotlib} \citep{Hunter:2007pv}, \texttt{Scikit-learn} \citep{scikit-learn}, \texttt{spaCy} \citep{spacy}}, \texttt{Fingerprint}.

\bibliography{strolger}{}
\bibliographystyle{aasjournal}



\end{document}